\documentclass[aps,prl,twocolumn,superscriptaddress,showpacs]{revtex4-1}

\usepackage{amsmath}
\usepackage{graphicx}
\usepackage{calc}
\usepackage{bm}

\usepackage[T2A]{fontenc}
\usepackage[cp1251]{inputenc}
\usepackage[english]{babel}

\begin{document}

\title{ Dynamic negative capacitance regime in GeTe Rashba ferroelectric.}

\author{N.N. Orlova}
\author{A.V. Timonina}
\author{N.N. Kolesnikov}
\author{E.V. Deviatov}

\affiliation{Institute of Solid State Physics of the Russian Academy of Sciences, Chernogolovka, Moscow District, 2 Academician Ossipyan str., 142432 Russia}

\date{\today}

\begin{abstract}
We experimentally investigate capacitance response of a thick ferroelectric GeTe single-crystal flake on the Si/SiO$_2$ substrate, where p-doped Si layer serves as a gate electrode. We confirm by resistance measurements, that for three-dimensional flakes,  electron concentration is not sensitive to the gate electric field due to the screening by bulk carriers. Unexpectedly, we observe that sample capacitance $C$ is strongly diminishing for both gate field polarities, so $C(V_g)$ is a maximum near the zero gate voltage. Also, we observe well-developed hysteresis with the gate voltage sweep direction for the experimental $C(V_g)$ curves. From our analysis, the capacitance behavior is explained by the known  dependence of the Rashba parameter on the  electric field for giant Rashba splitting in GeTe. In this case, the hysteresis in capacitance should be ascribed to polarization evolution in GeTe surface layers, which also allows to realize the regime of dynamic negative capacitance. The latter can be directly observed in time-dependent resistive measurements, as non-monotonic evolution of voltage response to the step-like current  pulse. Thus,  the  negative capacitance regime  can indeed improve performance and, therefore, the  energy efficiency of electronic devices.
\end{abstract}

\pacs{71.30.+h, 72.15.Rn, 73.43.Nq}

\maketitle

\section{Introduction}

Recent interest to systems with broken  inversion or time-reversal symmetries is mostly connected with topological materials. In Weyl semimetals (WSM), symmetry requires splitting of every Dirac point with three-dimensional linear spectrum into two Weyl nodes with different chiralities~\cite{armitage}. Fermi arcs are connecting projections of Weyl nodes on the surface Brillouin zone, which defines  topologically protected surface state transport in topological semimetals. On the other hand, the bulk properties are also affected by symmetry. In particular,  Weyl semimetals with broken time-reversal symmetry are bulk ferromagnetics or antiferromagnetics, while non-magnetic WSMs with broken inversion symmetry have to obtain ferroelectric polarization~\cite{armitage,TSreview}. In addition to  the gapless bulk spectrum~\cite{armitage}, ferroelectric polarization makes  non-magnetic Weyl semimetals being the natural  representation of the  novel concept of the  intrinsic polar metal, or the ferroelectric conductor~\cite{PM,pm1,pm2,pm3,pm4}. 

For ordinary ferroelectric insulators, the effect of  negative capacitance~\cite{NC1} is attracting significant interest nowadays~\cite{NC3, NC2}. Apart from the fundamental importance,  it is expected that the  negative capacitance field-effect transistor (NCFET)  can further improve the energy efficiency of electronic devices~\cite{NC2, NC4}. For a standard field-effect transistor, the capacitance between the gate and the quantum well can be written~\cite{shash,khrap} as 
\begin{equation}\label{C}  
     \frac{1}{C} = \frac{1}{C_0}+\frac{1}{Ae^2(dn_s/d\mu)},
\end{equation}	
where $C_0$ is the geometrical capacitance value, while the $Ae^2(dn_s/d\mu)$ term is the contribution from the finite thermodynamic density of states $dn_s/d\mu$ in the quantum well (the so called quantum capacitance), $A$ is the sample area. Since  $dn_s/d\mu$ is defined by the energy spectrum, capacitance spectroscopy is a powerful spectroscopic tool~\cite{Dolgop}, also for newest systems like   graphene~\cite{graph1,graph2,graph3,graph4} or topological insulators~\cite{TI1,TI2}.

It is accepted to speak about negative capacitance if the measured capacitance $C$ exceeds the geometrically defined value $C_0$ in Eq.(\ref{C}), which was firstly demonstrated for two-dimensional electron systems~\cite{Eisenst, dorozh} in GaAs  quantum wells. Nominally, negative capacitance requires  negative thermodynamic density of states $dn_s/d\mu$ in Eq.(\ref{C}), which is generally not allowed for macroscopically stable systems~\cite{NC1}. For the experimentally available systems, this regime  is stabilized by static external electric field, e.g. from ionized donors in quantum wells~\cite{Eisenst, dorozh,efros} or from another ferroelectric layer in NCFET~\cite{NC2, NC_SC}. Also, dynamic (transient) negative capacitance is proposed  due to the  polarization  switching in ferroelectrics~\cite{NC2, NC_SC}. Transient negative capacitance appears for  any microscopic switching mechanism as long as polarization  changes faster than charge. However, the device will always end up in a state of positive capacitance after the ferroelectric has switched~\cite{NC2, NC_SC}. 

Similar experiments can also be proposed for polar metals~\cite{PM,pm1,pm2,pm3,pm4}, or conductive ferroelectrics. In this case, screening of the gate electric field by bulk carriers allows to study only surface layer capacitance in a quasi-two-dimensional regime~\cite{Eisenst, dorozh}.  Among these materials, GeTe is of special interest~\cite{GeTespin-to-charge,GeTereview} due to the reported  giant Rashba splitting~\cite{GeTerashba}, which can be important also in spintronic applications. The ferroelectric control of spin-to-charge conversion was shown in epitaxial GeTe films~\cite{GeTespin-to-charge}, also, direct correlation between ferroelectricity and spin texture was demonstrated in this material~\cite{spin text}. Ferroelectric polarization is observed for the epitaxial films, microwires and bulk GeTe crystals~\cite{GeTespin-to-charge}, it is defined by the non-centrosymmetric distorted rhombohedral structure ($\alpha-GeTe$) with space group R3m (No. 160)~\cite{GeTerashba}.

Here, we experimentally investigate capacitance response of a thick ferroelectric GeTe single-crystal flake on the Si/SiO$_2$ substrate, where p-doped Si layer serves as a gate electrode. We confirm by resistance measurements, that for three-dimensional flakes,  electron concentration is not sensitive to the gate electric field due to the screening by bulk carriers. Unexpectedly,  we observe well-developed hysteresis with the gate voltage sweep direction for the experimental non-linear $C(V_g)$ curves. From our analysis,  the hysteresis in capacitance should be ascribed to polarization evolution in GeTe surface layers, which also allows to realize the regime of dynamic negative capacitance.

\section{Samples and techniques}

  The GeTe single crystals were grown by physical vapor transport in the evacuated silica ampule. The initial GeTe load was synthesized by direct reaction of the high-purity (99.9999\%) elements in vacuum. For the GeTe crystals growth, the initial load serves as the source of vapors: it was melted and kept at 770-780$^\circ$ C for 24 h. Afterward, the source was cooled down to 350$^\circ$ C at the 7.5 deg/h rate. The GeTe crystals grew during this process on the cold  ampule walls somewhat above the source. The GeTe composition is verified by energy-dispersive X-ray spectroscopy. The powder X-ray diffraction analysis confirms single-phase GeTe, the known structure model~\cite{GeTerashba} is also refined with single crystal X-ray diffraction measurements. 
 
 For the capacitance measurements we use thick (about 0.5~$\mu$m) GeTe single crystal flakes on the standard Si/SiO$_2$ substrate.  The p-doped Si layer serves as the gate electrode to apply the gate electric field  through the 200~nm thick SiO$_2$ layer, while conductive GeTe bulk allows nearly perfect field screening. 
 
 The $\approx$100~$\mu$m wide flakes are obtained  by mechanical exfoliation from the initial ingot. We choose the $\approx$100~$\mu$m wide flakes with defect-free surface by optical microscope. A single flake is placed by the most defect-free surface on Si/SiO$_2$ substrate  with pre-defined 10~$\mu$m wide Au contact leads. The leads pattern is prepared by lift-off technique after thermal evaporation of 100~nm Au. After initial single-shot pressing by another oxidized silicon substrate, no pressure is needed to hold the flake on the SiO$_2$ surface with Au leads.  
 
 This procedure with pre-defined contacts is specially designed for thick conductive flakes, where the desired experimental geometry can not be defined by usual mesa etching. It was demonstrated to  provide electrically and mechanically stable contacts with highly transparent metal-semiconductor interfaces and it also allows to avoid chemical or thermal treatment of the initial materials~\cite{black,SnSeour1,SnSeour2,WTeour,WTe2shapiro,WTe2chiral}.  Also, the relevant (bottom) GeTe surface is protected from any oxidation or contamination by a SiO2 substrate and thick  GeTe bulk. This is an important difference from  conventional monolayers, so we never observed surface degradation for the present sample design~\cite{black,SnSeour1,SnSeour2,WTeour,WTe2shapiro,WTe2chiral}.
 
Even for relatively thick flakes,  ferroelectric polarization is sensitive~\cite{WTeour} to the gate electric field, since the relevant (bottom) flake surface is directly adjoined to the SiO$_2$ layer.

To obtain $C(V_g)$ dependences of the sample capacitance on the dc gate voltage $V_g$, gate bias is applied between  the silicon substrate and one of the Au  leads, see the inset to Fig.~\ref{C(Vg)+circuit}. The dc bias is modulated by a low ac component. Both the imaginary and real ac current  components $Im I, Re I$ are measured in the circuit by lock-in. 

Au leads also allow resistance measurements for GeTe flakes. We obtain $dV/dI(I)$ or $dV/dI(V_g)$ curves in a two-point technique for the direct comparison with the two-point capacitance measurements: contact resistances always have some effect in capacitance measurements as non-zero real ac current component.   For the resistive measurements, the applied current $I$ is also modulated by low (0.1 mA) ac component, two-point ac voltage ($\sim dV/dI$) is obtained as a function of the dc source-drain current $I$ ($\pm 30$~mA) or the dc gate voltage $V_g$. 

We check directly by the gate electrometer that there is no measurable leakage current through the 200~nm thick SiO$_2$ layer at least in the  gate voltage range  $\pm 20$~V.  In the present setup, SiO$_2$ substrate protects the flake from further oxidation/contamination~\cite{black}, while the ferroelectric $\alpha-GeTe$  phase exists~\cite{Tc} below $700$~K. Thus, all the measurements are performed at room temperature under ambient conditions.

\section{Experimental results}

\begin{figure}
\center{\includegraphics[width=\columnwidth]{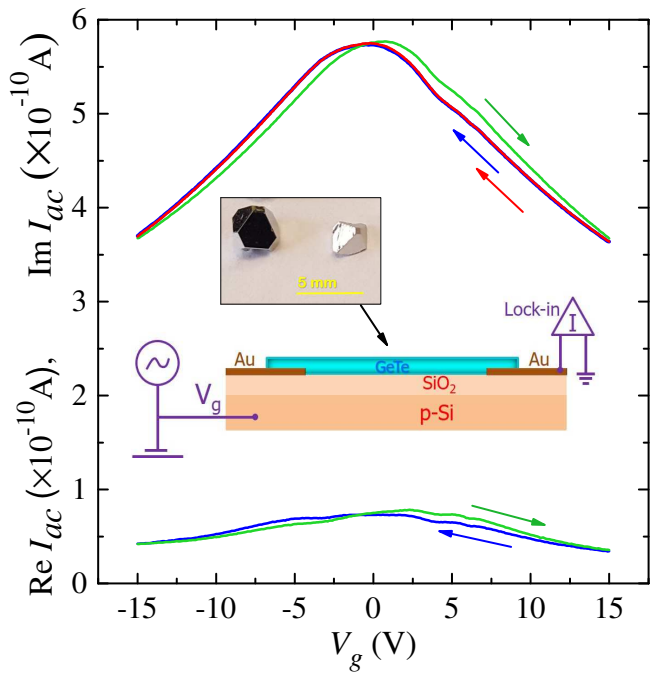}}
\caption{(Color online) $Im I, Re I$ ac current components for two opposite gate voltage sweep directions. The capacitance  $C \sim Im I$ dominates in the sample impedance because of $Im I >> Re I$ regime. In contrast with standard  field-effect devices~\cite{shash,khrap,dorozh}, $Im I_{ac}=2\pi fV_{ac} C$ is monotonously diminishing for both gate voltage polarities, so the experimental curves are nearly symmetric. Also, there is well-developed hysteresis with gate voltage sweep direction, as denoted by green and red arrows. On the other hand, experimental curves are well reproducible  for the same sweep direction (blue and red ones), so the sample is perfectly stable. 
The inset shows schematic diagram of the sample with electrical connections. Thick (about 0.5~$\mu$m) GeTe single crystal flake is obtained from the initial crystals (see the image). The flake is  placed   on the standard Si/SiO$_2$ substrate, so the p-doped bulk Si serves as the gate electrode to apply the gate bias. For the capacitance response, the dc bias is modulated by a low ac voltage, both the imaginary and real ac current  components are measured in the circuit by lock-in.}
\label{C(Vg)+circuit}
\end{figure}

Fig.~\ref{C(Vg)+circuit} shows both ac current components $Im I, Re I$ for two opposite gate voltage sweep directions. A typical gated sample can be described by RC circuit model, where $C$ is the gate capacitance while $R$ is the characteristic circuit resistance~\cite{shash,khrap,dorozh}. For the chosen ac modulation frequency $f$, we obtain  $Im I >> Re I$ in Fig.~\ref{C(Vg)+circuit} with strictly  linear $Im I (f)$ dependence   within 100 to 1000~Hz frequency  range. For this reason, the capacitance $C \sim Im I/f$ dominates in the sample impedance~\cite{shash,khrap,dorozh}. 

There are two nontrivial experimental observations in Fig.~\ref{C(Vg)+circuit} in comparison with  usual $C(V_g)$ curves for standard  field-effect devices. First, the sample capacitance $C \sim Im I$ is monotonously diminishing for both gate voltage polarities with a maximum around zero $V_g$, so the experimental curves are nearly symmetric.

As  the second  observation, Fig.~\ref{C(Vg)+circuit} shows well-developed hysteresis of the experimental curves with gate voltage sweep direction. It is important, that  two experimental curves are well reproducible  for the same sweep direction, as it is also depicted in  Fig.~\ref{C(Vg)+circuit}, so the sample is perfectly stable. The  hysteresis can not be expected for our electrical circuit, because of the negligible charging constant RC, which is well below $1/f$ for the $Im I >> Re I$ regime. The hysteresis can not be also connected with standard SiO$_2$ dielectric, which we have verified in the independent tests. Thus, the hysteresis in capacitance should be ascribed to the slow polarization evolution~\cite{SnSeour1,SnSeour2,WTeour} in GeTe surface layers. 

\begin{figure}
\center{\includegraphics[width=\columnwidth]{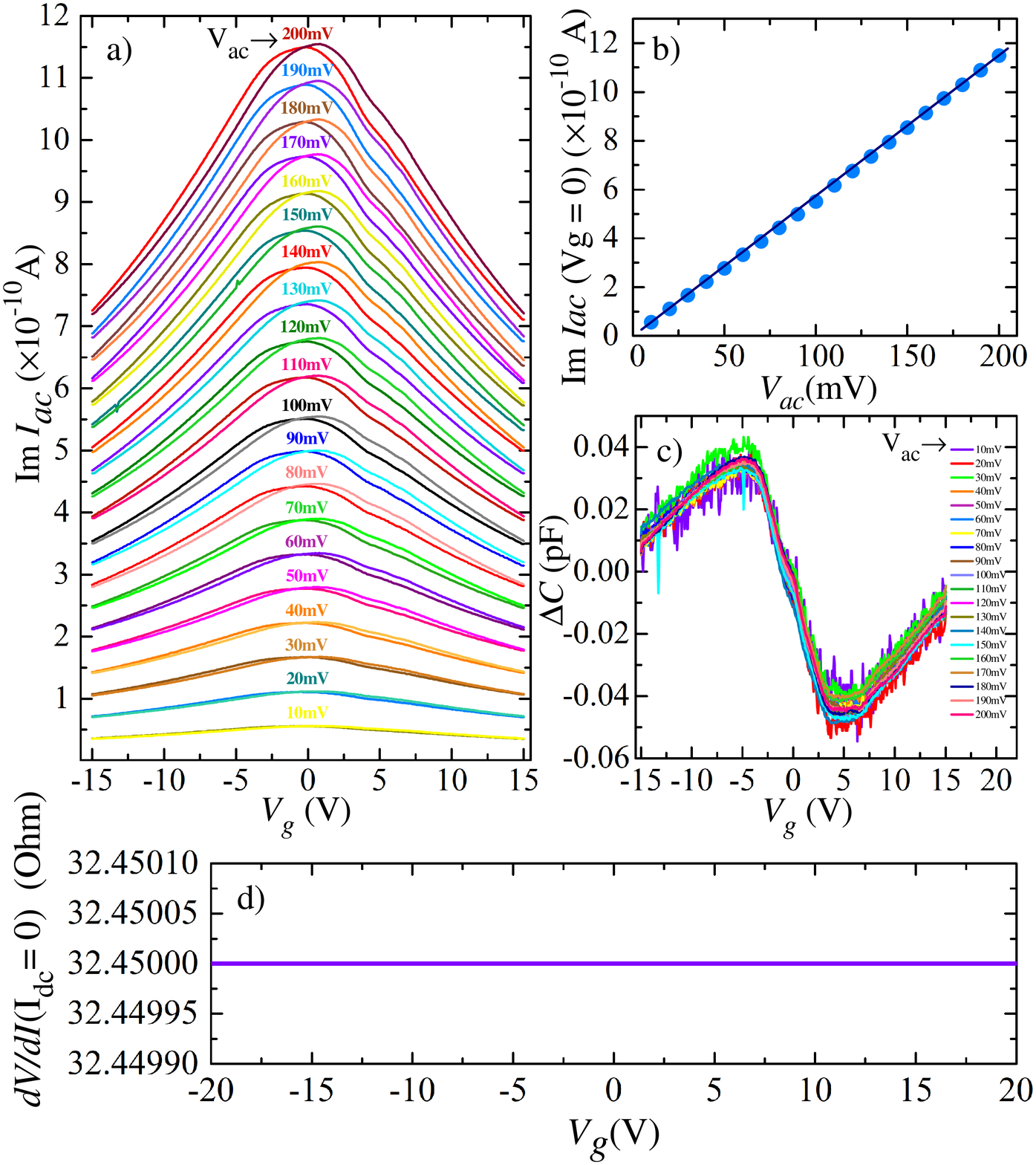}}
\caption{ (Color online) (a)  $Im I (V_g)$ curves for  different ac modulation voltage amplitudes $V_{ac}$  from 10 to 200~mV, the curves are shown for two opposite gate voltage sweep directions. (b) Linear dependence of  $Im I_{ac}(0)$  on  $V_{ac}$,  which satisfies the modulation-independent capacitance  $ C= Im I/2\pi fV_{ac}$ in Eq.~(\ref{C}). The obtained $C=5.8$~pF  gives the $\approx 10^4\mu$m$^2$ effective sample area, being quite reasonable for the 100~$\mu$m wide flakes over narrow Au contact leads.  (c) The capacitance difference $\Delta C$, obtained from  $\Delta Im I$ for the  difference between two $Im I (V_g)$ sweeps $\Delta Im I=Im I (V_g^+)-Im I (V_g^-)$.  From  the alternating $\Delta C$ sign, the measured capacitance exceeds the geometrically defined value for one of the remanent ferroelectric states. This behavior is known as the dynamic negative capacitance~\cite{NC2, NC_SC}. (d) Constant two-point  differential sample resistance $dV/dI$ in  a wide gate voltage range, so the carrier concentration in thick conductive GeTe flakes is non sensitive to the gate electric field.}
\label{Vac variation}
\end{figure}

Fig.~\ref{Vac variation} (a-b) confirms the independence of the measured capacitance on the ac modulation voltage amplitude.  Experimental $Im I_{ac}(V_g)$ curves are shown for two opposite sweep directions in Fig.~\ref{Vac variation} (a) for $V_{ac}$ values from 10 to 200~mV. At zero gate voltage, we demonstrate strictly linear dependence $Im I_{ac}(0) \sim V_{ac}$ in  Fig.~\ref{Vac variation} (b), as it should be expected in a linear regime $Im I_{ac}=2\pi fV_{ac} C$ of Eq.~(\ref{C}), the slope corresponds to $C=5.8$~pF value. The latter gives $\approx 10^4\mu m^2$ effective sample area, which is a reasonable estimation for the $\approx$100~$\mu$m wide GeTe flake.

For every modulation voltage $V_{ac}$, the hysteresis amplitude can be extracted as the difference between two sweeps $\Delta Im I=Im I (V_g^+)-Im I (V_g^-) $, 
 the result is shown in  Fig.~\ref{Vac variation} (c) as the capacitance difference $\Delta C=\Delta Im I/2\pi fV_{ac}$. All the $\Delta C$ curves are perfectly scaled into the  single dependence, so $\Delta C$ is only determined by the gate voltage $V_g$ sweep direction.  Also, $\Delta C (V_g)$ does not depend on the magnitude of the $V_g$ sweep, so $\Delta C$ is not determined by surface degradation. As an additional argument, the latter should affect the geometrical capacitance $C \sim 1/d$ through the effective thickness $d$, while we demonstrate perfectly linear dependence in Fig.~\ref{Vac variation} (b).

In contrast to the standard two-dimensional materials like graphene or field-effect transistors, carrier concentration in thick conductive GeTe flakes should not be sensitive to the gate electric field due to the screening of the gate electric field by bulk carriers. Fig.~\ref{Vac variation} (d) shows constant two-point  differential sample resistance $dV/dI$ in  wide gate voltage range. The measured resistance  $R=32.45$~Ohm allows also to estimate the RC circuit constant as $2\cdot10^{-10}$~s $<<1/f$. 

We wish to emphasize that, despite significant effect of the gate electric field on the capacitance (i.e. on the density of states in Eq.~(\ref{C})), it has no effect on the carrier concentration itself. This behavior also indicates the ferroelectric polarization effects~\cite{SnSeour1,SnSeour2,WTeour} in GeTe surface layers.

\begin{figure}
\center{\includegraphics[width=\columnwidth]{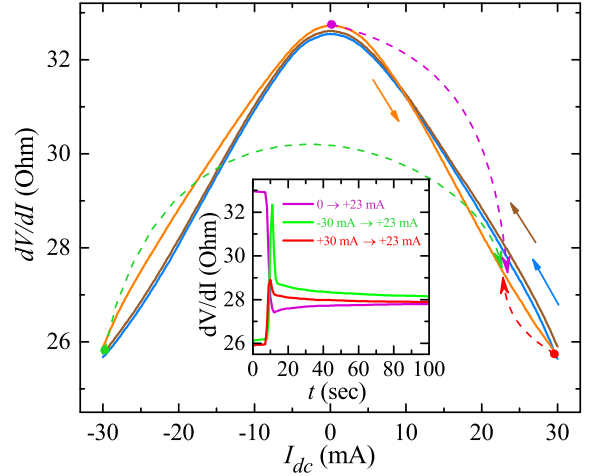}}
\caption{ (Color online) Polarization-current induced hysteresis with current sweep direction in two-point $dV/dI(I_{dc})$ curves, which is a known fingerprint of conductive ferroelectrics~\cite{WTeour,SnSeour1,SnSeour2}. Inset shows slow resistance relaxation  by time-dependent $dV/dI(t)$ curves. The sample resistance is stabilized for a long time at one of the dwelling bias values $I_{dc}=$+30~mA,-30~mA, or 0~mA. Afterward, the bias is abruptly changed to  $I_{dc}=+23$~mA and the time-dependent $dV/dI(t)$ relaxation is recorded for the 100~s interval. The relaxation is not monotonic: every $dV/dI(t)$ curve firstly overshoots the final $dV/dI$ value, and $dV/dI(t)$ slowly approaches to it afterward. This non-monotonic relaxation can not be expected for the trivial discharging $RC$-like process with positive, geometry-defined $C_0$ value, so it is a direct consequence of dynamically negative capacitance~\cite{NC2, NC_SC}.}
\label{dV/dI}
\end{figure}

For the two-point resistance measurements, non-linear $dV/dI(I_{dc})$ curves show hysteresis with current sweep direction in  Fig.~\ref{dV/dI}, which is a known fingerprint of conductive ferroelectrics~\cite{WTeour,SnSeour1,SnSeour2}. Current-induced source-drain field  shifts the ferroelectric domain walls in the sample, which results in additional polarization current. The latter  is connected with lattice deformation in ferroelectrics, and, therefore with slow resistance relaxation. The relaxation can be directly demonstrated by time-dependent $dV/dI(t)$ curves in the inset to Fig.~\ref{dV/dI}. For this curves, the sample resistance is stabilized for a long time at one of the distinct bias values $I_{dc}=$+30~mA,-30~mA, or 0~mA. Afterward, the bias is abruptly changed to  $I_{dc}=+23$~mA and the time-dependent $dV/dI(t)$ relaxation is recorded for the 100~s interval. The obtained $dV/dI(t)$ curves indeed show slow relaxation from different starting $dV/dI$ resistance values to the final one in the inset to Fig.~\ref{dV/dI}.  It seems to be important, that relaxation is not monotonic  in the inset to Fig.~\ref{dV/dI}: every $dV/dI(t)$ curve firstly overshoots the final $dV/dI$ value, and $dV/dI(t)$ slowly approaches to it afterward. This non-monotonic relaxation is unusual for the trivial discharging $RC$-like process, but can be expected for the dynamic (transient) negative capacitance in ferroelectrics~\cite{NC2, NC_SC}, so it can reflect the ferroelectric polarization dynamics in GeTe.  

\begin{figure}
\center{\includegraphics[width=\columnwidth]{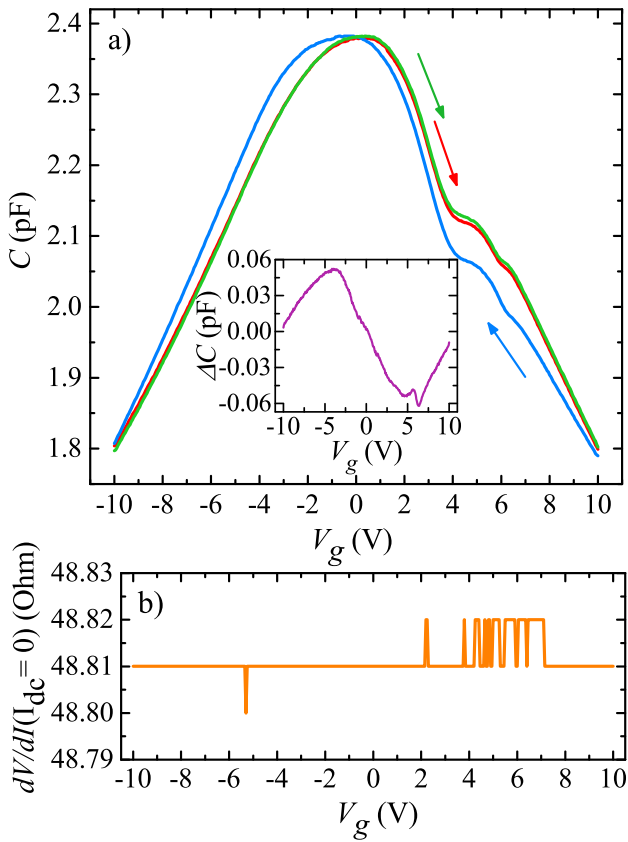}}
\caption{ (Color online) Hysteresis in $C(V_g)$ (a), the hysteresis amplitude $\Delta C$ of both signs (inset), and the gate-independent two-point  resistance $dV/dI(I_{dc}=0)$ (b), as obtained for a different sample. The above described scenario is well reproduced, so the dynamic negative capacitance due to the  polarization  switching~\cite{NC2, NC_SC} is inherent  for the gated GeTe structures. }
\label{diffr sample}
\end{figure}

The reported behavior can be well reproduced  for a different GeTe sample in Fig.~\ref{diffr sample}. We demonstrate the qualitatively similar results for the symmetric $C(V_g)$  dependence with hysteresis, the $\Delta C(V_g)$ hysteresis amplitude, and the gate-independent differential resistance $dV/dI$. One can also estimate $RC$ constant for this GeTe structure as $\approx 48.81\Omega \times 2.4\cdot10^{-12}$~F$\approx 1\cdot10^{-10}$~s from the absolute $R$ and $C$ values.

\section{Discussion}

Even qualitative $C(V_g)$ behavior in Figs.~\ref{C(Vg)+circuit} and~\ref{diffr sample} seems to be very unusual because of symmetric diminishing of the capacitance for both gate voltage polarities. 

For a two-dimensional electron gas in  standard field-effect transistors,  $dn_s/d\epsilon$ is a constant in zero magnetic field because of parabolic energy spectrum $\epsilon \sim p^2$ for electrons, so the capacitance is independent of the gate electric field~\cite{Dolgop,shash,khrap} in Eq.~(\ref{C}). For graphene samples with linear $\epsilon \sim p$, the capacitance~\cite{lozovik} grows  symmetrically  to both sides of the charge neutrality point~\cite{graph1,graph2,graph3,graph4}. Similar growth has been observed also for three-dimensional topological insulators~\cite{TI1,TI2}. In these structures, the capacitance reflects the charge accumulation at the interface (the so called field effect).

In our experiment, we observe symmetric diminishing of the capacitance in Figs.~\ref{C(Vg)+circuit} and~\ref{diffr sample}, while  there is no any effect of the gate electric field on the carrier concentration in Figs.~\ref{Vac variation} (d) and~\ref{diffr sample} (b).  The latter still can be estimated in the capacitor approximation as $\delta n / n \sim \delta R /R \approx 2\times10^{-4} $ for the sample from Fig.~\ref{diffr sample} (b). This ratio is quite low due to the high carrier concentration in three-dimensional  GeTe ( $\sim 10^{20}$cm$^{-3}$) in comparison with two-dimensional systems like monolayers and quantum wells ($\sim 10^{11}$cm$^{-2}$ for the typical 10~nm thickness). In a capacitor, the concentration variation is the primary effect of the gate voltage, despite some effect on the mobility can not be also excluded. 

In contrast to the mentioned above structures, GeTe is characterized by giant Rashba spin-orbit coupling~\cite{GeTereview}, so the Rashba term $\pm \alpha_Rk$ is dominant in density of states.
Since the carrier concentration is independent of the gate electric field in our samples (see Figs.~\ref{Vac variation} (d) and~\ref{diffr sample} (b)), it is the Rashba parameter $\alpha_R$ which defines the $C(V_g)$ behavior in Figs.~\ref{C(Vg)+circuit} and~\ref{diffr sample}. The dependence of the Rashba parameter on the ferroelectric polarization is known for giant Rashba splitting in GeTe from theoretical~\cite{GeTerashba} and experimental~\cite{spin text} investigations. Even full reversal of the Rashba parameter can be achieved upon reversal of the electric field~\cite{GeTerashba}.

The total device capacitance $C$ should be written as
\begin{equation}\label{CR}  
     \frac{1}{C} = \frac{1}{C_0}+\frac{1}{C_Q}. 
\end{equation}
The first term is a constant (geometrical capacitance), while the second (quantum capacitance) should be responsible for the gate voltage dependence.  The total $C$ value is diminishing, if the second term ($1/C_Q$) is increasing in Eq,(\ref{CR}) increase with gate voltage.  Since we are sure, that carrier concentration is independent on $V_g$ (see above), we can write the quantum capacitance for the dominant Rashba term  as 
$$
C_Q=Ae^2\frac{\sqrt{2\pi n_s}}{\pi \alpha_R \hbar},
$$
which gives required $1/C_Q\sim \alpha_R$ for the known gate-field dependent giant Rashba splitting in GeTe~\cite{GeTerashba}. For the significant effect, the maximum value of this quantum capacitance should be of the  same order  as the geometrical one. The latter is estimated above from the experimental data as $C_0\sim 1$~nF, as well as the effective sample area $A \approx 10^4\mu m^2$.  The sheet density $n_s$ can be estimated as $\sim 10^{14}$cm$^{-2}$ from the GeTe bulk value ($\sim 10^{20}$cm$^{-3}$) for the 10~nm effective layer thickness. For these parameters, the $C_Q\sim C_0$ criterion is achieved for the known $\alpha_R  \approx 30$~eV\AA      ~for GeTe~\cite{GeTerashba}. This estimation confirms our conclusion on the $C(V_g)$ diminishing in Figs.~\ref{C(Vg)+circuit} and~\ref{diffr sample}.

Noteworthy, that Ref.\cite{GeTerashba} connects the Rashba effect and the hysteretic nature of ferroelectricity.  In our experiment, it appears as two different values of capacitance obtained at every $V_g$ for two opposite sweep directions, i.e. for two different ferroelectric polarizations. In other words, the non-zero $\Delta C$ in Fig.~\ref{Vac variation} (c) indicates, that the measured capacitance exceeds the geometrically expected value for one of the remanent ferroelectric states.  The  remnant polarization can be estimated from the distance between two maxima in $C(V_g)$ curves as $E\sim V_g/d \sim 10^{7}$~V/m, where $d$ is the 200~nm SiO$_2$ layer thickness. Indeed, $E$ is the field, that shifts the  $C(V_g)$ maximum from the zero gate voltage for the symmetric $C(V_g)$ curve. In contrast, $\Delta C$ value is not so  straightforward for the analysis if the sample contains multiple ferroelectric domains~\cite{WTe2_fer}.  This behavior is known as the dynamic negative capacitance due to the  polarization  switching~\cite{NC2, NC_SC}, so the non-zero $\Delta C$ demonstrates it for the gated GeTe structure in Figs.~\ref{C(Vg)+circuit} and~\ref{diffr sample}.

The effect of dynamic negative capacitance can be directly seen in the time-dependent resistance measurements in Fig.~\ref{dV/dI} as non-monotonic voltage response to the step-like current  pulse. In general,  the response should monotonically saturate with $RC$ time constant for the geometry-defined $C_0$ value. In contrast, $dV/dI(t)$ curve firstly overshoots the final $dV/dI$ value in the inset to Fig.~\ref{dV/dI}, which is a direct result of dynamically negative $C$. Thus,  the negative capacitance regime can indeed improve performance and, therefore, the  energy efficiency of gated electronic devices~\cite{NC2}.

\section{Conclusion}
As a result, we experimentally investigate capacitance response of a thick ferroelectric GeTe single-crystal flake on the Si/SiO$_2$ substrate, where p-doped Si layer serves as a gate electrode. We confirm by resistance measurements, that for three-dimensional flakes,  electron concentration is not sensitive to the gate electric field due to the screening by bulk carriers. Unexpectedly, we observe that sample capacitance $C$ is strongly diminishing for both gate field polarities, so $C(V_g)$ is a maximum near the zero gate voltage. Also, we observe well-developed hysteresis with the gate voltage sweep direction for the experimental $C(V_g)$ curves. From our analysis, the capacitance behavior is explained by the known  dependence of the Rashba parameter on the  electric field for giant Rashba splitting in GeTe. In this case, the hysteresis in capacitance should be ascribed to polarization evolution in GeTe surface layers, which also allows to realize the regime of dynamic negative capacitance. The latter can be directly observed in time-dependent resistive measurements, as non-monotonic evolution of voltage response to the step-like current  pulse. Thus,  the  negative capacitance regime  can indeed improve performance and, therefore, the  energy efficiency of electronic devices.

\section{Acknowledgement}
We wish to thank S.S~Khasanov for X-ray sample characterization.

\end{document}